# Technical recommendation on multiplex MR elastography for tomographic mapping of abdominal stiffness with a focus on the pancreas and pancreatic ductal adenocarcinoma

**Authors:** Jakob Schattenfroh[1], Salma Almutawakel[1], Jan Bieling[1], Johannes Castelein[2] PhD, Melanie Estrella[1] PhD, Philippe Garteiser[3] PhD, Viktor Hartung[4] MD, Karl H. Hillebrandt[5] MD, Adrian T. Huber[6,7,8,9] MD, Laura Körner[10], Thomas Kröncke[11] MD, Thomas Malinka[5] MD, Hans-Jonas Meyer[12] MD, Tom Meyer[1], Uwe Pelzer[13] MD, Felix Pfister[12], Igor M. Sauer[5] MD, Anna Speth[1], Bernard E. Van Beers[3,14] PhD, Carsten Warmuth[1], PhD, Nienke P. M. Wassenaar[15,16], Yanglei Wu[1], Rolf Otto Reiter[1,17]* MD, Ingolf Sack[1]* PhD, * Equally contributing senior authors

**Corresponding Author:** Jakob Schattenfroh; jakob.schattenfroh@charite.de

On behalf of the European Consortium of MRE in PDAC

1.  Department of Radiology, Charité – Universitätsmedizin Berlin, Germany
2.  Department of Radiology, University Medical Center Gdańsk, Poland
3.  Center of Research on Inflammation, Université Paris Cité, Inserm, Paris, France
4.  Department of Diagnostic and Interventional Radiology, University Hospital Würzburg, Germany
5.  Department of Surgery, Charité – Universitätsmedizin Berlin, Germany
6.  Department of Radiology, Clinic Beau-Site, Hirslanden Group, Switzerland
7.  Radiology and Nuclear Medicine, Lucerne Cantonal Hospital, University of Lucerne, Switzerland
8.  Liver Elastography Center, Translational Imaging Center, Swiss Institute for Translational and Entrepreneurial Medicine, Bern, Switzerland
9.  Department of Diagnostic, Interventional and Pediatric Radiology, Inselspital Bern, University Hospital, University of Bern, Switzerland
10. Medical Physics in Radiology, German Cancer Research Center, Heidelberg, Germany
11. Department of Diagnostic and Interventional Radiology, University Hospital Augsburg, Germany
12. Department of Diagnostic and Interventional Radiology, Universitätsmedizin Leipzig, Germany
13. Medical Clinic for Haematology, Oncology and Cancer Immunology, Charité – Universitätsmedizin Berlin, Germany
14. Department of Radiology, Beaujon University Hospital Paris Nord, AP-HP, Clichy, France
15. Department of Radiology and Nuclear Medicine, Amsterdam University Medical Center, The Netherlands
16. Imaging and Biomarkers, Cancer Center Amsterdam, The Netherlands
17. Berlin Institute of Health at Charité – Universitätsmedizin Berlin, BIH Biomedical Innovation Academy, BIH Charité Digital Clinician Scientist Program, Germany

# Abstract

**Objectives:** MR elastography (MRE) offers valuable mechanical tissue characterization, however, in deep abdominal organs like the pancreas conventional single-driver, single-frequency approaches often fail. This study evaluates whether multiplex MRE using multiple drivers and vibration frequencies can overcome these limitations.

**Materials and Methods:** This prospective study used single-shot spin-echo MRE in 18 healthy volunteers (mean age 30 ± 8 years) targeting the liver, pancreas, kidneys, and spleen. Each healthy volunteer underwent 16 MRE examinations with different sets of four vibration frequencies in the range of 30-60 Hz and four driver combinations, and an additional null experiment without vibrations. Further, a cohort of 14 patients with pancreatic ductal adenocarcinoma (PDAC, mean age 57 ± 15 years) were retrospectively assessed. The quality of shear-wave fields and stiffness maps was assessed by displacement amplitudes and image sharpness.

**Results:** In healthy volunteers, abdominal coverage with displacement amplitudes above the pre-determined noise level of 4 µm varied between MRE configurations: 24.2% ([0.0% - 56.2%], single-driver at 60 Hz), 66.9% ([24.8% - 97.7%], single-driver at 30-60 Hz), 70.2% ([0.0% - 92.5%], multi-driver at 60 Hz) and 99.9% ([89.4% - 100%], multi-driver at 30-60 Hz). In the pancreas, more than 60% coverage was achieved in all subjects using four drivers and multiple frequencies. This was achieved in only 2 of 18 subjects (11%) using single-driver/single-frequency MRE. Superficial organs were adequately assessed in all configurations. Patients with PDAC had 99.1% [91.4% - 100%] coverage in the pancreas and 96.3% [63.1% - 100%] abdominal coverage (multi-driver at 30-60 Hz) suggesting the clinical feasibility of tomographic stiffness mapping by multiplex MRE.

**Conclusion:** MRE with at least four drivers and multiple vibration frequencies between 30-60 Hz enables tomographic mapping of tissue stiffness across the entire abdomen, including the pancreas. Multiplex MRE offers a promising approach for generating detailed images of abdominal stiffness, potentially enhancing clinical diagnostics for abdominal and pancreatic diseases.

# Introduction

The viscoelastic properties of soft tissues provide valuable diagnostic information that can be exploited by MR elastography (MRE)[1] to detect diseases such as liver fibrosis[2], chronic kidney disease[3], and pancreatic ductal adenocarcinoma (PDAC)[4]. Systemic conditions like hypertension can modify stiffness across organ boundaries, including hepatic, splanchnic, and renal hyperelastic stiffening[5,6]. In PDAC, the liver is the primary target for metastatic colonization and may be mechanically compromised before harboring metastatic cells[7]. These examples highlight the potential of tomographic assessments of mechanical properties across abdominal tissue and organ boundaries within a single MRE examination. Many MRE studies have published spatially resolved stiffness maps of abdominal organs[8]. However, these maps often lack complete coverage of the entire field-of-view (FoV), particularly in deep-seated and mechanically shielded tissues like the pancreas[9,10], limiting simultaneous multi-organ biomechanical assessment. The technical requirements to set up MRE for stiffness mapping throughout the entire abdomen remain undefined.

We hypothesize that MRE using shear-wave excitation at multiple frequencies with multiple drivers, henceforth referred to as *multiplex MRE*, improves whole-abdomen stiffness mapping, particularly in pancreatic tissue, compared to single-driver, single-frequency MRE[11]. To test this hypothesis, we prospectively enrolled healthy volunteers in a study of different combinations of vibration frequencies and independent drivers. The clinical applicability of the developed MRE quality metrics were retrospectively analyzed in a multiplex MRE cohort of patients with pancreatic ductal adenocarcinoma (PDAC).

Extending previous knowledge on the effectiveness of multi-driver shear-wave excitation in MRE of the human liver[12], this study aims to explore the limits of tomographic MRE in other organs and within a band of excitation frequencies toward routine clinical mapping of organ stiffness throughout the abdomen with a focus on deep-seated tissue such as the pancreas.

## Materials and Methods

*Study Design*

This prospective study was approved by the institutional ethics board. All patients and healthy volunteers provided written and verbal informed consent. Eighteen healthy volunteers (mean age 30 ± 8 years, 15 men, BMI 23.2 ± 2.2 kg/m$^2$) without any known abdominal diseases were recruited. All volunteers fasted ≥3 hours before examination (Figure 1A). To investigate possible influences of PDAC and non-tumorous pancreatic tissue on shear wave coverage, fourteen patients (mean age 57 ± 15 years, 5 men, BMI 23.4 ± 4.2 kg/m$^2$) with biopsy confirmed PDAC were randomly selected from a cohort of previously published multifrequency, multi-driver MRE[13].

*MRE Setup and Acquisition*

All healthy volunteers were examined using a 3-Tesla MRI scanner (Magnetom Lumina; Siemens Healthineers, Germany) and a spin-echo, echo-planar imaging sequence equipped with flow-compensated motion-encoding gradients (MEG)[14,15].

The MRE imaging parameters are listed in Table 1. Fifteen transverse contiguous slices were acquired during free breathing[16,17]. Four independent pressurized-air powered drivers, placed around the chest (Figure 1B), were used to excite mechanical vibrations in the liver, pancreas, kidneys, and spleen. The number of active drivers was incrementally increased from one to four (Figure 1C). Sixteen independent MRE acquisitions were performed per subject. Vibration frequencies were evaluated in four sets (60 Hz alone; 50 and 60 Hz combined; 40, 50 and 60 Hz combined; 30, 40, 50 and 60 Hz combined), each tested with one, two, three and four active drivers, respectively. Additionally, a null experiment without vibrations was performed to establish baseline noise levels. Patients with PDAC were investigated with an MRE setup similar to the multiplex study in healthy volunteers using four pressurized-air drivers and four driver frequencies of 30-60 Hz. Details of MRI scanners and sequences are provided in[13].

*Image Post-Processing*

Motion correction was performed by applying the realign method of the SPM12 toolbox[18] to the complex MRI data before MRE post-processing[16]. Maps of shear-wave speed (SWS, m/s) were reconstructed using the publicly available multi-frequency inversion method, k-MDEV, optimized for abdominal MRE (bioqic-apps.charite.de)[19,20]. 3D organ-specific regions of interest (ROIs) were drawn by a trained imaging physicist (>3 years of experience in abdominal MRE) based on the MRE magnitude images (Figure 2). To minimize boundary effects from major blood vessels and slip interfaces, voxels with frequency-averaged SWS values below 1 m/s were excluded[21].

*Quality Metrics*

To assess MRE performance, we defined four quality metrics for shear-wave fields and SWS maps:

(1) **Harmonic oscillatory amplitudes and the coefficient of variation (CV)** were used to measure wave displacement amplitudes and variability within the entire abdominal ROI as well as in specific organ regions. Displacement amplitudes (µm) were determined as the total amplitudes of complex wave images (root mean square of all MEG directions) at the fundamental vibration frequency and averaged over all applied frequencies.

(2) **Wave displacement threshold** was based on the probability density function (PDF) of displacement amplitudes of (1) with and without external vibrations to identify areas with insufficient wave displacement. The distributions of displacement amplitudes in each voxel of the ROI and averaged over all applied frequencies were compared between vibrations 'on' and 'off' conditions. Considering residual 'off' amplitudes as noise, only 'on' displacement amplitudes exceeding this noise-threshold[22] were further considered as relevant wave intensities. Wave coverage was quantified as the percentage of ROI voxels exceeding the displacement threshold.

(3) **Image sharpness of SWS maps** was determined in the full abdominal ROI as a measure of anatomical contrast across abdominal organs and structures. The method relied on the variance of the Laplacian operator applied to the SWS maps[16].

(4) **MRE success rates in specific organs** were determined using the wave displacement threshold defined in (2). Success rates, expressed as percentages of studied subjects, were determined based on whether pre-determined thresholds of 25%, 60%, or 95% of all voxels within an organ exhibited sufficiently high wave amplitudes.

*Statistical Analysis*

All data were statistically analyzed using MATLAB (R2024b, The MathWorks), with parametric tests (ANOVA, t-tests, correlation coefficient), two-way ANOVA to assess effects of number of drivers and frequencies on coverage, displacement, and SWS. Inter-subject confidence interval (CI) was 95% with a significance level of $p < 0.05$.

# Results

*Wave displacement amplitudes with four drivers*

The PDFs of harmonic oscillation amplitudes based on the four frequencies from 30 to 60 Hz and four drivers are shown in Figure 3. The pooled amplitudes over the entire abdominal ROI, averaged over four driver frequencies, were used to delineate the threshold of shear-wave amplitudes corresponding to the PDF intercept of amplitudes with and without vibrations at 4.0 ± 0.3 µm. The mode (peak in the shear-wave amplitude histogram) for the vibration 'on' condition was 7.0 ± 0.9 µm, whereas the residual amplitudes in the vibration 'off' condition, treated as background noise, were 1.1 ± 0.1 µm (Figure 3A). The frequency-resolved analysis (Figure 3B) revealed that amplitudes increased with decreasing frequency. With a mode of 15.5 ± 1.8 µm and IQR of 2.7 ± 0.3 µm, intra-abdominal displacement amplitudes were largest at 30 Hz, while at higher frequencies, mean amplitudes and IQR decreased with -0.38 µm/Hz (r = -0.85, p < 0.001) and -0.06 µm/Hz, (r = -0.62, p < 0.001), respectively. Region-specific analysis compounding four excitation frequencies of 30-60 Hz showed similar mean amplitudes of 6.5 ± 1.5 µm, 9.7 ± 2.0 µm, and 10.1 ± 1.8 µm in the liver, kidney and spleen, respectively, whereas lower amplitudes of 5.8 ± 1.1 µm were found in the pancreas (ANOVA: p < 0.001, Figure 3C). Mode and IQR values deduced from Figure 3 are summarized in Table 2.

*Number of excitation frequencies and active drivers*

Figure 4 illustrates the effect of driver setup and excitation frequency on the shear-wave field, wave displacement amplitude and SWS maps. The numbers of active drivers and excitation frequency increased wave displacement amplitudes throughout the abdominal region (p < 0.001). Increasing the number of active drivers enhanced the uniformity of displacement amplitudes (p < 0.001) as indicated by a more homogeneous wave propagation and reduced spatial variation in shear-waves in Figure 4A. Figure 4B demarcates regions with displacement amplitudes <4 µm, showing areas of insufficient displacement amplitude predominantly at higher frequencies and fewer drivers. Figure 4C quantifies the percentage of voxels exceeding the 4 µm amplitude threshold. Multi-driver, multi-frequency setups increased displacement amplitudes and wave coverage of the abdominal region from a median of 24.2% (range, 0.0% - 56.2%) with single-driver, 60 Hz MRE to a median of 99.9% (range, 89.4% - 100%) with four drivers, 30-60 Hz MRE (p < 0.001). Increasing the number of drivers from one to four increased the coverage irrespective of vibration frequency (p < 0.001). Furthermore, larger ranges indicated greater intersubject variability in coverage at higher frequencies and with fewer drivers. Figure 4D shows SWS maps generated by 16 different setups. In regions with displacement amplitudes of <4 µm SWS maps were considered unreliable and masked out, particularly relevant for deep tissues such as the pancreas. While deep tissues require four drivers and multi-frequency excitation, unaffected superficial areas as in the liver and spleen can be found using single-driver MRE at 60 Hz.

These observations were quantitatively corroborated by the group statistical plots of four frequencies and variable driver counts in Figure 5. Wave displacement amplitudes increased (r = 0.74, p < 0.001) and CV decreased (r = -0.77, p < 0.001) with number of drivers, indicating improved wave coverage and homogeneity (Figures 5A-B). Figure 5C presents the impact of the individual frequencies on SWS sharpness (r = 0.75, p < 0.001), based on four-driver acquisition, suggesting enhanced overall anatomical resolution and SWS map quality achieved by lower frequencies. Both the number of active drivers (ANOVA: p = 0.02) and frequencies (ANOVA: p < 0.001) positively influenced image sharpness.

*MRE success rates*

Figure 6 shows organ-specific success rates defined by percentage wave coverage. Specifically, considering MRE to be feasible if >60% of an organ is covered by wave amplitudes >4 µm, liver MRE at 60 Hz based on a single driver could be performed in 4 of 18 volunteers (22%). However, this success rate decreased to 2 of 18 volunteers (11%) in the pancreas, whereas multiplex MRE based on four drivers and four frequencies was successful in all volunteers. Similar analyses based on 25% and 95% wave coverage are (Supplemental Figure S1-2) suggest that the tomographic display of SWS including nearly all (>95%) pancreatic voxels was feasible in 15 of 18 volunteers if based on four drivers and four frequencies. In contrast, single-driver 60 Hz MRE was successful in 9 of 18 subjects, limited to superficial liver tissue, when applying a more relaxed criterion of 25% wave coverage. Group mean SWS values of all organs and settings are listed in supplemental Table S1.

*Multiplex MRE in PDAC*

In patients with PDAC, consistent with observations made in healthy volunteers using the same multi-driver, multi-frequency setup, 99.1% [91.4% - 100%] coverage of the pancreas and 96.3% [63.1% - 100%] coverage of the whole abdomen with shear waves was found. The pre-specified quality threshold of >60% organ coverage with displacement amplitudes >4 µm was achieved in all patients for both non-tumorous pancreatic tissue and PDAC tissue subregions, demonstrating the robustness of multiplex MRE for tomographic stiffness imaging in a clinical setting. Also wave displacement amplitudes and variability were similar to those in healthy volunteers (mean displacement: 5.7 ± 2.7 µm, CV: 0.46 ± 0.14). Non-tumorous pancreatic tissue had a mean value of 5.7 ± 2.9 µm, while PDAC tissue subregions showed a slightly lower mean of 4.4 ± 2.2 µm. Despite tumor-related tissue heterogeneity, PDAC tissue exhibited consistent wave propagation, with a CV of 0.26 ± 0.08. In contrast, non-tumorous pancreatic tissue exhibited a slightly higher CV of 0.37 ± 0.08, probably due to the larger volume.

## Discussion

The alteration of abdominal tissue stiffness across organ boundaries due to systemic diseases motivates full-FoV stiffness mapping by tomographic MRE. Complementing many multi-organ analyses in MRE literature, this is, to the best of our knowledge, the first systematic report on technical requirements of multi-driver, multi-frequency (multiplex) wave excitation for tomographic MRE of the abdomen.

The systemic biomechanical characterization and establishment of reference values of healthy tissue of the liver (SWS = 1.36 ± 0.10 m/s), kidney (SWS = 2.06 ± 0.29 m/s), spleen (SWS = 2.08 ± 0.32 m/s), and pancreas (SWS = 1.31 ± 0.12 m/s) may become increasingly relevant for the detection of pressure-related diseases [23], tissue matrix alterations [24], cancer development [25] or treatment response [26]. Our results encourage the setup of multiplex MRE with four drivers and four vibration frequencies. Operating multiple parallel drivers is inexpensive while measurement time increases with the number of frequencies applied. The ability to perform scans in under 3 minutes under free breathing while incorporating four frequencies, indicates that multiplex MRE can be seamlessly integrated into routine clinical abdominal MRI protocols.

Our findings in the PDAC cohort validate the benefit of multiplex MRE observed in healthy volunteers, particularly for imaging deep-seated pancreatic tissue. The ability to reliably measure stiffness in PDAC and adjacent non-tumorous tissue suggests multiplex MRE could aid in early diagnosis, tumor characterization, monitoring disease progression, and potentially evaluating treatment response in clinical settings.

The study by Chen et al.[12] investigated single-frequency (60 Hz) MRE in the liver, using multiple passive pneumatic drivers, each powered by an active driver. Assessing MRE map quality with a semi-quantitative reliability criterion based on subjective wave propagation scores, the authors concluded that multi-driver MRE improves shear-wave coverage across the abdomen, but did not offer a practical technical solution, merely noting technical challenges of managing multiple active drivers. In contrast, the compressed air setup used in this study facilitates multi-driver MRE by using parallel pressure tubes from a single air outlet to power additional drivers, allowing easy scaling to more drivers.

It is a limitation of this study that we did not perform such experiments with more than four drivers and frequencies. However, it should be noted that each volunteer underwent a total of 17 independent MRE examinations (4 frequency settings x 4 drivers + 1 control experiment). This extensive protocol, applied to 18 healthy participants, provided the statistical power necessary to verify our hypothesis and to recommend the technical setup of multiplex MRE as a tomographic modality in the abdomen and pancreas. Further tests in patients with altered abdominal stiffness and larger BMI are planned.

In summary, this study analyzed the MRE coverage of abdominal organs with one to four independent compressed-air drivers in a range of frequencies between 30 to 60 Hz toward tomographic MRE of the pancreas. It was found that only multiplex MRE based on four drivers and four vibration frequencies could successfully investigate pancreatic tissue in all subjects while superficial tissues could be assessed by standard single-driver MRE at 60 Hz. Given a scan time of under 3 min it is recommended to use multiplex MRE when a tomographic display of stiffness in abdominal organs, particularly of the pancreas, is desired. This recommendation was retrospectively validated in multiplex MRE cohort of patients with PDAC

## Acknowledgements

The authors gratefully acknowledge financial support from the German Research Foundation (DFG 540759292 (M5) to R.R. and I.S., DFG FOR5628 to R.R. and I.S., SFB1340 to I.S., and BIOQIC GRK 2260 to I.S.).

**Table 1:** Imaging parameters for abdominal multiplex MRE at 3-Tesla MRI

| Parameter | Value |
| --- | --- |
| Coil | Tx: body coil, Rx: 12-channel surface coil, 24-channel spine-array |
| Repetition time, echo time | 1540 ms, 50 ms |
| Voxel size | 2.5×2.5×5.0 mm$^3$ |
| Field of view | 280×350×75 mm$^3$ |
| Matrix | 112×140×15 |
| Fat suppression | SPAIR (spectrally selective adiabatic inversion recovery) |
| Bandwidth | 1552 Hz/Px |
| Acceleration | GRAPPA 2, Partial Fourier 7/8 |
| MEG amplitude, duration | 34 mT/m, 17.6 ms |
| MEG encoding efficiency for vibration frequencies in brackets | 47 (30 Hz), 72 (40 Hz), 93 (50 Hz), 105 (60 Hz) rad/mm |
| MRE vibration phase offsets | 8 |
| Acquisition time for 3 MEG directions and 4 frequencies | 2:40 min |

**Table 2: *Statistical parameters of the PDF curves shown in Figure 3.*** *Mode and interquartile range (IQR) with 95% CI for different conditions: Vibration 'on' and 'off', different frequencies, and different organs.*

|  | vib 'on' | vib 'off' | 30 Hz | 40 Hz | 50 Hz | 60 Hz | Liver | Kidney | Spleen | Pancreas |
|---|---|---|---|---|---|---|---|---|---|---|
| **Mode, µm** | 7.0±0.9 | 1.1±0.1 | 15.5±1.9 | 9.8±1.0 | 6.7±0.8 | 4.6±0.6 | 6.5±1.5 | 9.7±2.0 | 10.1±1.8 | 5.8±1.1 |
| **IQR, µm** | 4.1±0.3 | 0.2±0.2 | 2.7±0.2 | 3.1±0.2 | 2.5±0.2 | 1.4±0.3 | 3.1±0.6 | 4.1±0.6 | 3.5±0.6 | 4.5±0.7 |

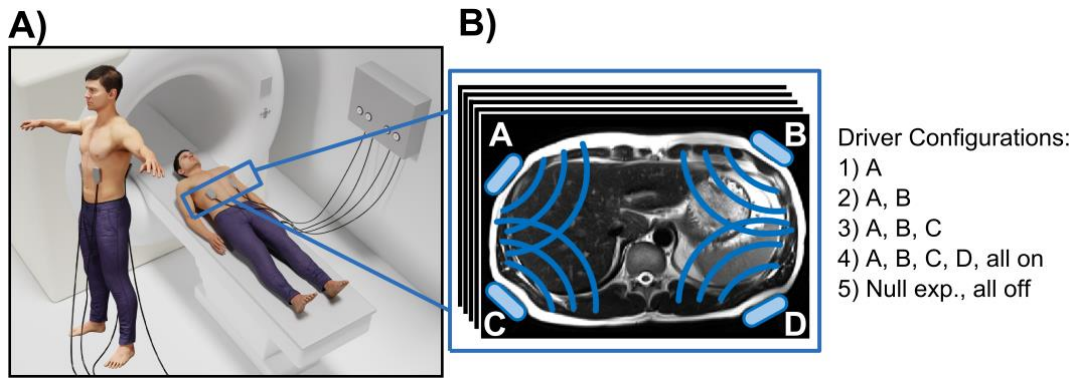

**Figure 1: *Technical setup. A)*** *Experimental setup, and **B)** and driver placement with four drivers (A-D) generating shear-waves. Driver configurations from one active driver to four active drivers and a null experiment were investigated.*

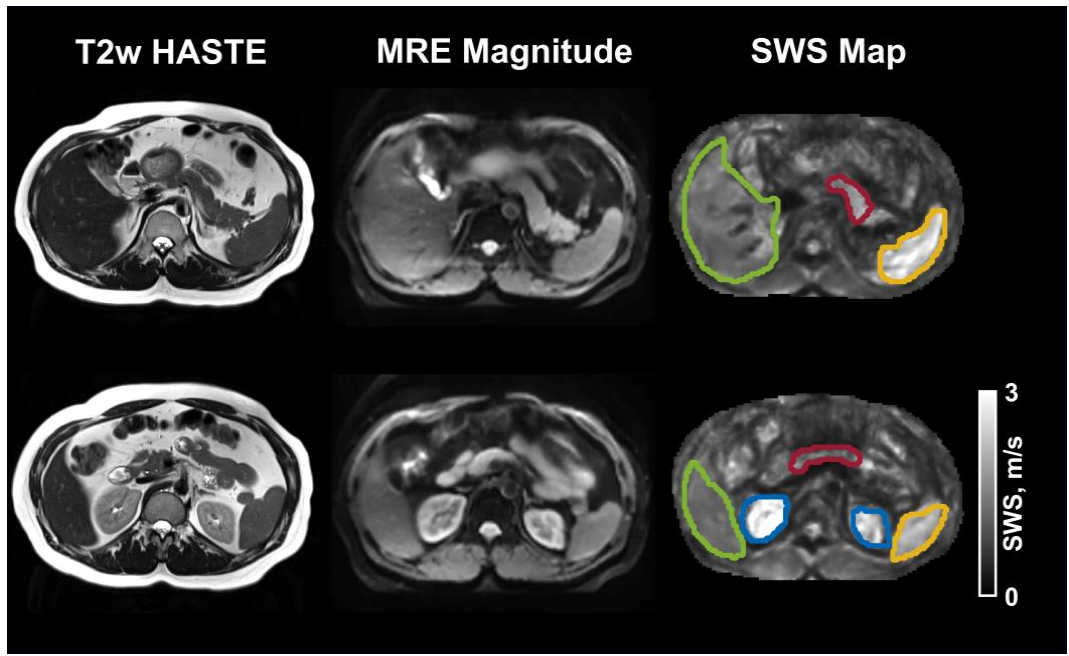

**Figure 2:** *Two transversal MRI and MRE images in a healthy volunteer.* Anatomical reference images (T2-weighted HASTE), MRE magnitude images, and stiffness maps given as shear-wave speed (SWS) with organ-specific ROIs (green: liver, yellow: spleen, blue: kidney, red: pancreas) of two slices of a representative subject.

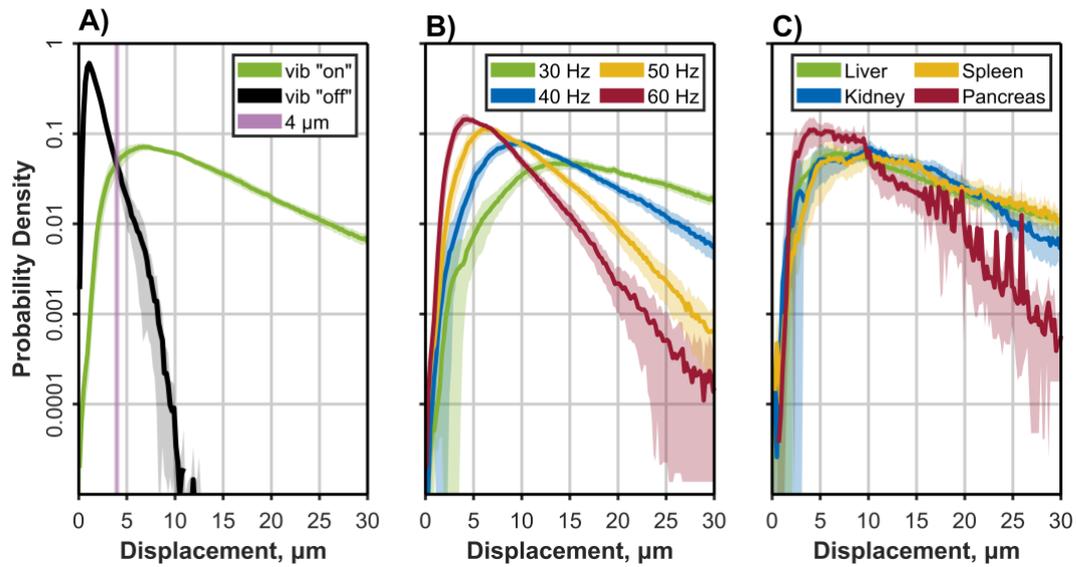

**Figure 3:** *Probability density functions (PDFs) of oscillatory amplitudes (displacement) in different regions of interest (ROI), at different frequencies and based on four external drivers.* **A)** *Oscillatory amplitudes in the full abdominal ROI, averaged over four frequencies from 30 to 60 Hz with and without vibrations ('on' - green, 'off' - black). The vertical purple line demarcates the amplitude threshold of 4 µm, corresponding to the PDF-intersection, above which oscillatory amplitudes were taken as MRE shear-waves, while below this threshold, amplitudes were deemed insufficient or noise.* **B)** *Frequency-resolved PDF within the abdominal ROI, showing higher oscillatory amplitudes at lower frequencies. Combining the four frequencies in multiplex MRE results in higher averaged amplitudes throughout the abdominal organs, as shown in C).* **C)** *Organ-specific PDF averaged over four frequencies from 30 to 60 Hz, showing lowest amplitudes in the mechanically shielded pancreatic region. Shaded areas represent intersubject CI of 95%.*

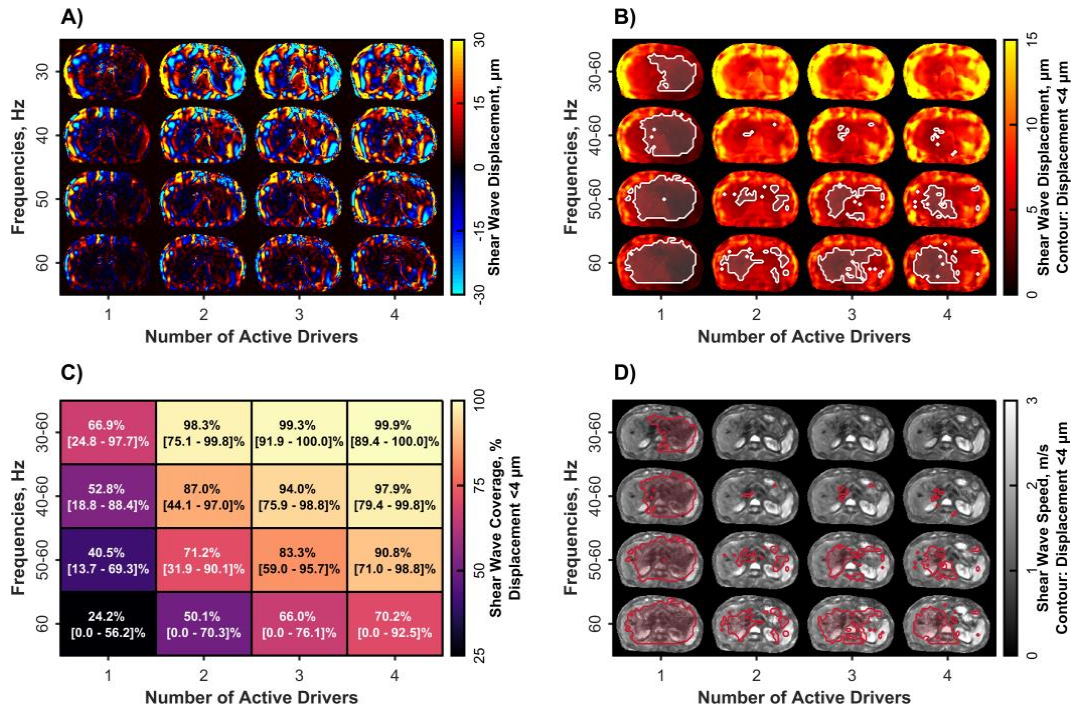

**Figure 4:** *Effects of driver count and excitation frequency on shear-wave displacement and coverage of abdominal SWS maps. **A)** Shear-wave maps for different excitation frequencies and varying numbers of active drivers, demonstrating increased displacement amplitudes with more drivers and lower vibration frequencies. **B)** Maps of the shear-wave amplitude with contours indicating displacement <4 μm, highlighting improved wave propagation with more drivers and frequencies. **C)** Median shear-wave coverage and range (percentage of valid voxels with amplitudes >4 μm over total abdominal ROI volume) indicating increased coverage with both more active drivers and including more excitation frequencies. **D)** SWS maps across all conditions with insufficient wave displacement (<4 μm amplitudes) demarcated in red.*

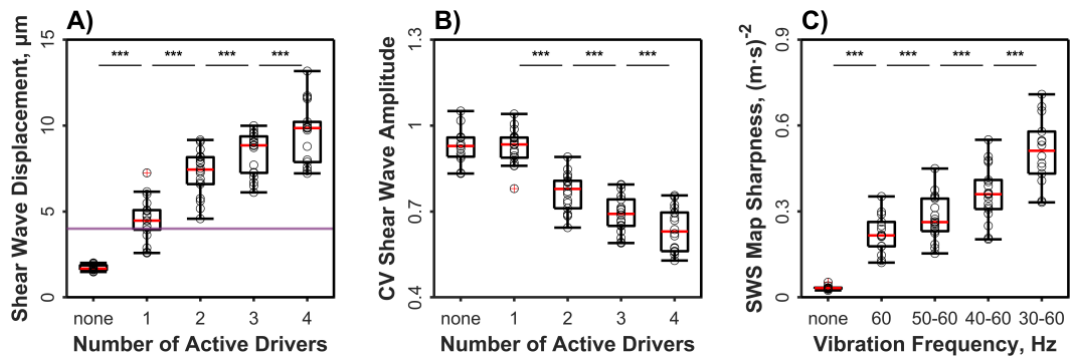

**Figure 5:** *Full abdominal ROI: group statistics of shear-wave displacement and SWS sharpness. **A)** Oscillatory amplitudes for four compounded frequencies (combined 30-60 Hz) over number of drivers. The horizonal line indicates the 4 µm noise threshold. **B)** Coefficient of variation (CV) of shear-wave amplitudes (combined 30-60 Hz) showing their increasing homogeneity within the abdominal ROI by adding more drivers. **C)** SWS map sharpness over vibration frequency bins, indicating increased detail resolution when including lower vibration frequencies.*

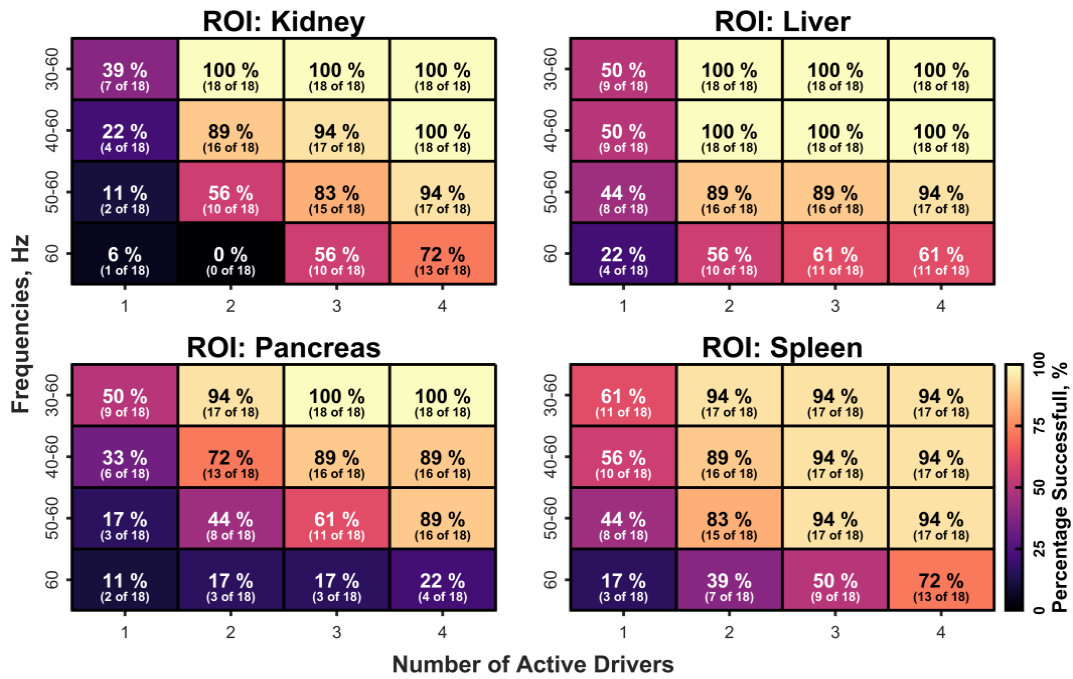

**Figure 6:** *MRE success rates in abdominal organs. MRE examinations were considered successful if >60% of organ-specific ROI showed wave amplitudes above the noise level of 4 μm.* Colors and numbers show the rate of successful MRE in our cohort given this 60% threshold. A similar figure is presented as supplemental material for 25% and 95% thresholds (Figures S1 and S2).

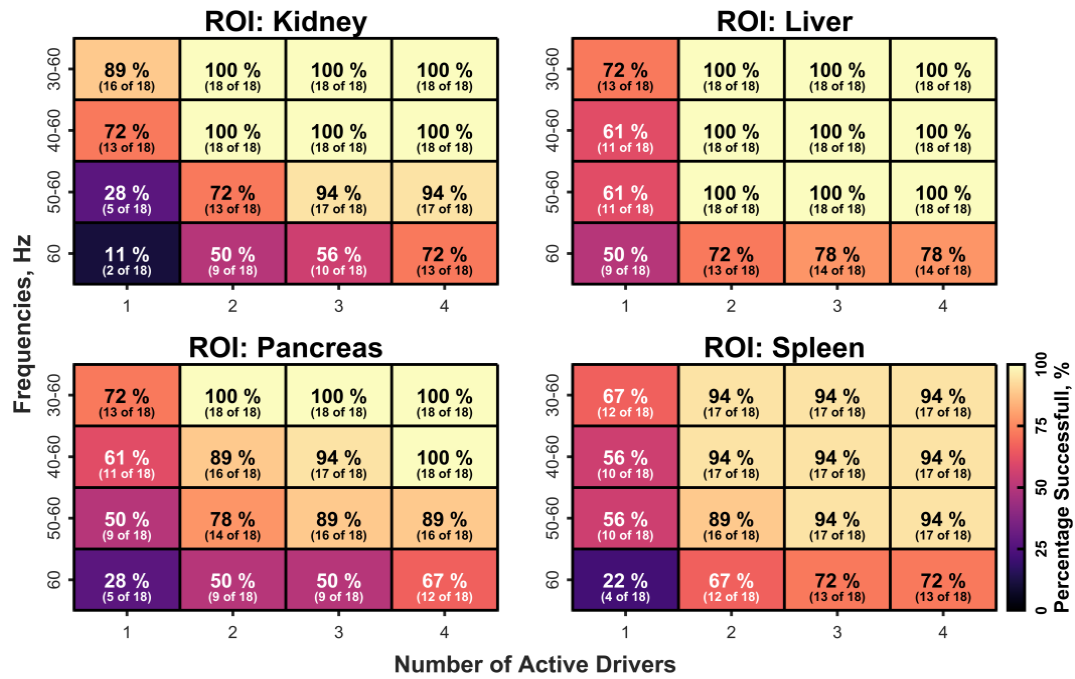

**Figure S1:** MRE success rates in abdominal organs at a 25% threshold (>25% ROI with wave amplitudes >4 μm).

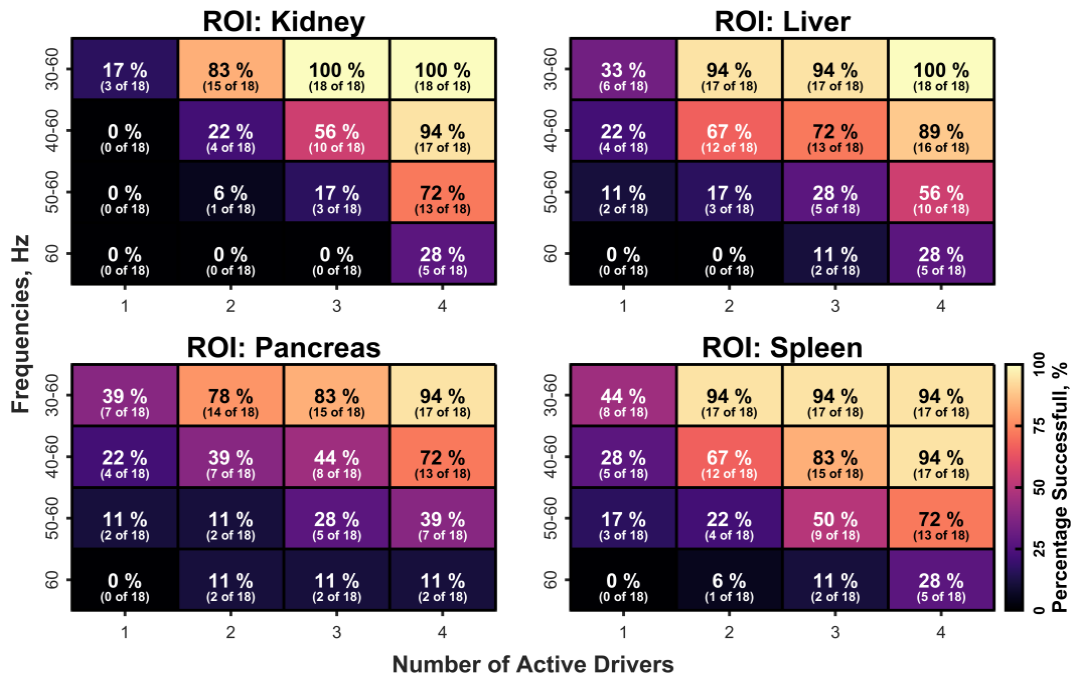

**Figure S2:** MRE success rates in abdominal organs at a 95% threshold (>95% ROI with wave amplitudes >4 μm).

| Organ | Frequencies, Hz | Mean Shear Wave Speed (SWS), m/s | | | |
|---|---|---|---|---|---|
| | | Number of Active Drivers | | | |
| | | 1 | 2 | 3 | 4 |
| Kidney | 30 - 60 | 1.65±0.22 | 1.83±0.27 | 1.97±0.31 | 2.06±0.29 |
| | 40 - 60 | 1.86±0.27 | 1.97±0.36 | 2.12±0.40 | 2.24±0.37 |
| | 50 - 60 | 2.07±0.44 | 2.10±0.47 | 2.25±0.53 | 2.31±0.55 |
| | 60 | 2.30±0.27 | 2.53±0.32 | 2.81±0.19 | 2.77±0.30 |
| Liver | 30 - 60 | 1.33±0.10 | 1.35±0.10 | 1.35±0.11 | 1.36±0.10 |
| | 40 - 60 | 1.46±0.15 | 1.40±0.13 | 1.40±0.14 | 1.41±0.13 |
| | 50 - 60 | 1.55±0.17 | 1.45±0.16 | 1.44±0.18 | 1.45±0.17 |
| | 60 | 1.70±0.13 | 1.54±0.25 | 1.59±0.13 | 1.51±0.25 |
| Spleen | 30 - 60 | 1.87±0.29 | 2.01±0.29 | 2.05±0.32 | 2.08±0.32 |
| | 40 - 60 | 1.98±0.31 | 2.11±0.34 | 2.16±0.34 | 2.19±0.32 |
| | 50 - 60 | 2.03±0.37 | 2.19±0.44 | 2.24±0.41 | 2.27±0.39 |
| | 60 | 2.34±0.29 | 2.53±0.40 | 2.54±0.33 | 2.55±0.32 |
| Pancreas | 30 - 60 | 1.27±0.18 | 1.30±0.14 | 1.32±0.13 | 1.31±0.12 |
| | 40 - 60 | 1.44±0.26 | 1.43±0.20 | 1.44±0.19 | 1.42±0.14 |
| | 50 - 60 | 1.71±0.44 | 1.56±0.29 | 1.61±0.33 | 1.51±0.20 |
| | 60 | 2.04±0.64 | 1.78±0.32 | 1.83±0.52 | 1.75±0.31 |

**Table S1:** Mean SWS values (± standard deviation) for the kidney, liver, spleen, and pancreas for the 16 combinations of frequency ranges (30-60 Hz, 40-60 Hz, 50-60 Hz, 60 Hz) and number of active drivers (1 to 4).